\documentclass[aps,prl,twocolumn,groupedaddress,showpacs]{revtex4}

\usepackage{graphicx}
\usepackage{color}
\usepackage{epsfig}
\usepackage{amsmath}

\begin{document}

\title{Signals of the QCD phase transition in core-collapse supernovae}

\author{I. Sagert,
M. Hempel,
G. Pagliara
and
J. Schaffner-Bielich$^\star$}
\affiliation{
\begin{tabular}{c}
Institute for Theoretical Physics, Goethe University, 
Max-von-Laue-Str.~1, 60438 Frankfurt am Main, Germany\\
$\star$Institute for Theoretical Physics, Ruprecht-Karls-University, Philosophenweg 16, D-69120 Heidelberg, Germany 
\end{tabular}
}

\author{T. Fischer$^\dag$,
A. Mezzacappa$^\ddag$,
F.-K. Thielemann$^\dag$
and
M. Liebend\"orfer$^\dag$}
\affiliation{
\begin{tabular}{c}
  \dag  Department of Physics, University of Basel, Klingelbergstr.~82, 4056 Basel,           
                      Switzerland \\
  \ddag Physics Division, Oak Ridge National Laboratory, Oak Ridge, TN~37831
\end{tabular}
}

\date{\today}

\begin{abstract}
  We explore the implications of the QCD phase transition during the
  postbounce evolution of core-collapse supernovae.
  Using the MIT bag model for the description of quark matter 
  and assuming small bag constants, we find that the
  phase transition occurs during the early postbounce accretion phase.
  This stage of the evolution can be simulated with general
  relativistic three-flavor Boltzmann neutrino transport.
  The phase transition produces a second shock wave that triggers  a
  delayed supernova explosion. 
  If such a phase transition happens in a future
  galactic supernova, its existence and properties should become 
  observable as a second peak in the neutrino signal 
  that is accompanied by significant
  changes in the energy of the emitted neutrinos. In contrast to the first neutronization
burst, this second neutrino burst is dominated by the emission of
anti-neutrinos because the electron-degeneracy is lifted when the
second shock passes through the previously neutronized matter.
\end{abstract}

\pacs{
12.38.-t 
12.38.Mh 
21.65.Qr 
25.75.Nq  
26.30.-k 
26.50.+x 
26.60.-c 
26.60.Kp 
95.30.Sf 
95.55.Vj 
95.85.Ry 
97.60.Bw 
}
\maketitle
In search of the phase transition from hadronic to deconfined 
matter, heavy ion experiments at RHIC and at LHC 
at CERN explore the QCD phase diagram for large
temperatures and small baryochemical potentials. For these conditions, 
which were also present in the early universe, 
lattice QCD calculations predict a crossover transition 
between the deconfined chirally symmetric phase and the 
confined phase with broken chiral symmetry. For high chemical 
potentials and low temperatures a first order chiral 
phase transition is expected and will be tested at the FAIR facility at GSI Darmstadt.

Due to their large central densities, compact stars can also
serve as laboratories for nuclear matter beyond saturation density
and may contain quark matter \cite{Weber:2005}. 
The formation of quark matter in compact stars is mainly discussed in
two scenarios, in protoneutron stars (PNS) after the supernova explosion
\cite{Pons:2001ar} and in old accreting neutron stars
\cite{Lin:2005zda,Abdikamalov:2008df}.
For the first case, deleptonization leads to the loss of lepton
pressure and therefore to an increase in the central density so that the
phase transition takes place. Possible observables are the emission of gravitational waves
\cite{Lin:2005zda,Abdikamalov:2008df} due to the contraction 
of the neutron star or delayed $\gamma$-ray bursts 
\cite{Berezhiani:2002ks}.

In this article we want to follow a third and less discussed case.  
The phase transition from hadronic to quark matter 
can already occur  in the 
early postbounce phase of a core-collapse supernova 
\cite{TakaharaSato:1988a,TakaharaSato:1988b,
Gentile:1993ma,Drago:1997tn,Yasutake:2007st}. 
This requires a phase transition onset close to saturation density, 
which can be realized for high temperatures and low proton fractions. 
For such a scenario Ref.~\cite{Gentile:1993ma} 
found the formation of a second shock
as a direct consequence of the  phase transition.
However, the lack of neutrino transport in their model
allowed them to investigate the dynamics 
only for a few ms after bounce.
Very recently, a quark matter phase transition has
been considered with Boltzmann neutrino transport 
for a 100 M\(_\odot\) progenitor \cite{Nakazato:2008su}.
The appearance of quark
matter shortened the time until black
hole formation due to the softening of the equation of state (EoS), 
but did not lead to the launch of a second shock.

In our core-collapse simulations
of low- and intermediate-mass progenitor stars,
we confirm the formation of a second shock
caused by the phase transition to quark matter.
We even find that the second shock
triggers a delayed supernova explosion during 
the postbounce accretion phase.
This represents an interesting addition 
to currently discussed supernova
explosion mechanisms, such as the neutrino-driven
\cite{BetheWilson:1985},
the magneto-rotational \cite{LeBlankWilson:1970,Bisnovatyi-Kogan:1971}
or the acoustic mechanisms \cite{Burrows:etal:2006}. 
A clear imprint of the phase transition to quark matter
at the launch of the explosion could be expected in the
neutrino signal, the emission of gravitational waves and
the nucleosynthesis yields. Given the improvements in the sensitivity of neutrino detectors since
SN1987A, it seems that the neutrino signal from a Galactic supernova is
the most likely observable which could provide a first indication of the
early phase transition to quark matter.

\begin{figure}
\includegraphics[width=\columnwidth]{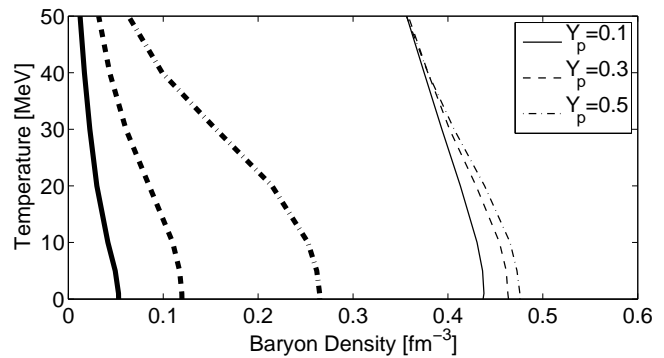}
\caption{\label{fig:critdens} The onset (thick lines) and the end (thin lines) of the
  mixed phase from the QCD phase transition from nuclear matter to quark
  matter using \emph{eos1}.}
\label{fig1}
\end{figure}

Lattice QCD is not yet able to make predictions for
large chemical potentials relevant for neutron star calculations.
Consequently, the quark matter EoS is currently computed
using phenomenological descriptions such as the MIT bag
or the NJL models. 
As we aim to study the basic 
effects from quark matter phase transitions on 
core-collapse supernovae, we chose the 
very simple but widely applied MIT bag model. 
In modeling the phase transition to quark matter 
there is a main physical uncertainty: the critical density \(n_{crit}\)
for the onset of the mixed phase. In our model,
\(n_{crit}\) is determined by the 
bag constant \(B\) and the strange quark mass.
At present, the bag constant is not a fixed parameter 
and is usually chosen between $B^{1/4}=145-200$ MeV \cite{Schertler:2000xq}. We choose the bag constant such that
we obtain an early onset for the phase transition and a maximum
mass of more than 1.44 M$_\odot$, without enabling absolutely stable
strange quark matter. Within this narrow range we select $B^{1/4}=162$
MeV (\emph{eos1}) and 165 MeV (\emph{eos2}), and a strange quark mass of 100 MeV
as indicated by the Particle Data Group \cite{Eidelman:2004wy}. The two choices of
the bag constants lead to critical densities of $\sim0.12$ fm$^{-3}$ and $\sim 0.16$ fm$^{-3}$, respectively (for $T=0$ and proton fraction $Y_p=0.3$). For the hadronic EoS we use the table of Shen et al. 
\cite{Shen:1998gq}. The phase transition to quark matter is 
modeled by a Gibbs construction as discussed in 
Refs.~\cite{Drago:1997tn,Nakazato:2008su} and is therefore of second order according to the Ehrenfest
classification but a phase transition of the first kind after Landau's
definition \cite{1969stph.book.....L}. The phase diagram using \emph{eos1} for different proton fractions \(Y_p\) 
is shown in Fig.~\ref{fig:critdens}.
We remark that larger values for the bag constant result in higher critical densities.
For the sake of simplicity, we have 
neglected finite size effects and Coulomb interactions 
within the mixed phase. Such contributions would decrease 
the density range of the mixed phase making the phase
transition more similar to a Maxwell construction 
\cite{Voskresensky:2002hu}. 
For comparison, we have also computed the EoS 
by using the Maxwell construction and find that the results of our core-collapse
simulations are qualitatively similar.

We note that the small obtained values for the critical density 
close to saturation density are not in contradiction with heavy ion data. In contrast to heavy-ion collisions high-density supernova matter is isospin-asymmetric with a proton fraction \(Y_p\sim 0.3\). Furthermore SN timescales are in the range of ms and therefore long enough to establish equilibrium with respect to weak interactions that change the strangeness on the timescales of micro-seconds or less. The additional strange quark degree of freedom and the large asymmetry energy
allow one to obtain small values for \(n_{crit}\) and lead to an early
appearance of quark matter (see Fig.~\ref{fig:critdens}). 

Our choice of parameters is also compatible
with the still most precise neutron star mass measurement 
of \(1.44\) M\(_{\odot}\) for the Hulse-Taylor pulsar \cite{LattimerPrakash:2004}. 
With \emph{eos1} and \emph{eos2}, we obtain values for the 
maximum gravitational mass of 1.56 and 1.50 M\(_{\odot}\) respectively.
Note that higher neutron star masses can be achieved with
more sophisticated models of quark matter \cite{Alford:2006vz}.
For \(B^{1/4}\)=162 and 165 MeV, 
almost the entire star is composed of quark 
matter, surrounded by a mixed quark-hadronic phase, 
which is enclosed by a thin pure hadronic crust. 

For the accurate prediction of the three-flavor 
neutrino signal, general relativistic effects may be important.
Hence, we choose for our investigation a well-tested general relativistic description of the neutrino 
radiation hydrodynamics in spherical symmetry, that
is based on Boltzmann neutrino transport
\cite{MezzacappaBruenn:1993c,
MezzacappaMesser:1999,
Liebendoerfer:etal:2004}.
Quark matter appears only in optically thick regimes where
neutrinos are in thermal and chemical equilibrium with matter. For this first proof-of-principle study, we use hadronic weak
interaction rates \cite{Bruenn:1985} in the quark phase and derive
the hadronic chemical potentials from the quark chemical potentials
such that weak equilibrium for neutrinos in quark matter is obtained.
Our simulations are launched from a
10 and a 15 M\(_\odot\) progenitor model
from Ref.~\cite{Woosley:etal:2002}.

\begin{figure}
  \centering
    \includegraphics[width=\columnwidth]{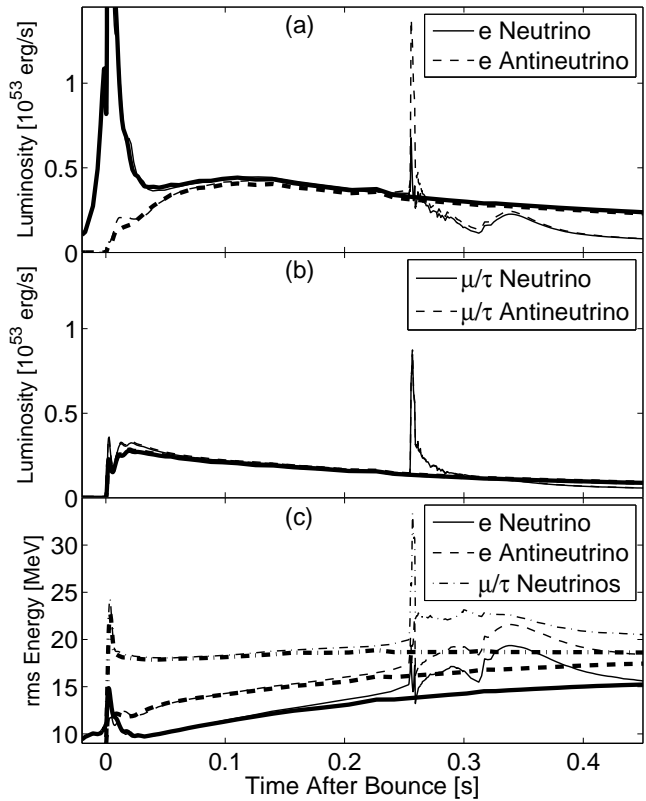}
  \caption{\label{fig:luminosity-energy}
            Neutrino luminosities (a) and (b)
            and  rms-energies (c) measured at 500 km distance
	    for a 10 M\(_\odot\) progenitor model.
	    The results of the quark EoS \emph{eos1} (thin lines) are compared to
	    the results of the pure hadronic EoS \cite{Shen:1998gq} (thick lines).
	    A second neutrino burst is clearly visible at \(\sim\) 260 ms after bounce.}
\end{figure}

The standard core-collapse scenario
leads to core bounce at nuclear saturation density
and the formation of a shock.
This expanding shock looses energy
due to the dissociation of nuclei and the emission
of the \(\nu_e\)-burst at \(\sim\) 10 ms after bounce 
(see Fig.~\ref{fig:luminosity-energy} (a))
and therefore turns into a standing accretion shock (SAS).
The SAS could be revived by neutrino heating \cite{BetheWilson:1985}.
However, explosions in spherically symmetric models with
accurate neutrino transport have only been obtained
for a 8 M\(_\odot\) ONeMg progenitor star \cite{Kitaura:etal:2006}.
The collapse of more massive progenitors leads to
an extended postbounce phase, during which
the central PNS contracts due to mass
accretion. 

\begin{table}
\begin{center}
\begin{tabular}[b]{ccccccc}
\hline 
\hline
Prog. 
& 
EoS 
& 
\(t_{pb}\) 
& 
M\(_{Q}\) 
& 
M\(_{mixed}\) 
& 
M\(_{PNS}\) 
& 
E\(_{expl}\)\\

[M\(_\odot\)]
&

&
[ms]
&
[M\(_\odot\)]
&
[M\(_\odot\)]
&
[M\(_\odot\)]
&
[10\(^{51}\)erg]
\tabularnewline 
\hline 
10
&
\emph{eos1}
&
255
&
0.850
&
0.508
&
1.440
&
0.44
\tabularnewline
10
&
\emph{eos2}
&
448
&
1.198
&
0.161
&
1.478
&
1.64
\tabularnewline
15
&
\emph{eos1}
& 
209
&
1.146
&
0.320
&
1.608
&
0.42
\tabularnewline
15
&
\emph{eos2}
&
330\footnote{moment of black hole formation}
&
1.496
&
0.116
&
1.700
&
\mbox{unknown}\footnote{black hole formation before positive explosion energy is achieved}
\tabularnewline
\hline
\hline
\end{tabular}
\caption{Selected properties of the different models.
\(t_{pb}\) is the time after bounce at the first appearance 
of the pure quark phase.
The different mass distributions (baryon mass) of the PNS, 
the quark mass M\(_Q\), the mass of the mixed phase M\(_{mixed}\)
and the total PNS mass M\(_{PNS}\) are taken
at a later stage when clearly positive explosion energies E\(_{expl}\)
are obtained.}
\label{table}
\end{center}
\end{table}

In our models that allow a transition to quark matter, 
the onset of the mixed phase
(\(n_B \sim\) 0.1 fm\(^{-3}\), \(T \simeq 10\) MeV,
\(Y_e \sim\) 0.3)
is already achieved at core bounce.
The initially reached quark matter 
fraction at the center of the PNS 
remains small during the first 50 ms after bounce. 
In the subsequent compression,
the quark matter fraction rises again 
and an increasing central region of the PNS 
enters the mixed phase. The reduced adiabatic index 
causes the PNS to collapse.

\begin{figure}
  \centering
  \includegraphics[width=\columnwidth]{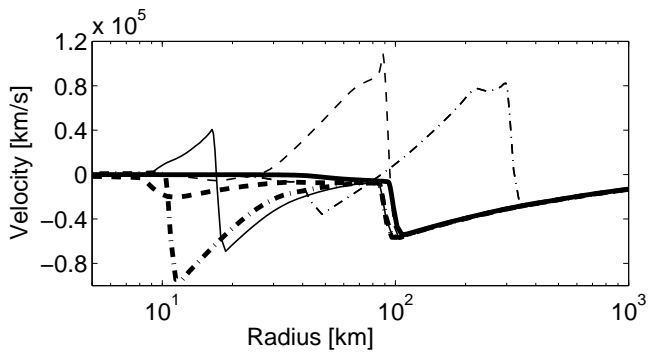}
  \caption{\label{fig:hydroplot}
           Velocity profiles at different times during the postbounce evolution
           of a 10 M\(_\odot\) progenitor model based on \emph{eos1},
           illustrating the development of the explosion through different stages;
           thick solid line at 240.5 ms, 
	   dashed at 255.2 ms, 
	   dash-dotted at 255.5 ms
           and 
	   thin solid line at 255.6 ms, 
	   dashed at 256.4 ms, 
	   dash-dotted at 258.3 ms.}
\end{figure}

\emph{PNS collapse:} The central density rises up to
4 - 5 times nuclear saturation density where the collapse halts 
due to the stiffening of the EoS in the pure quark phase. 
A large fraction of the PNS is composed of quarks,
enclosed by a mixed hadronic-quark phase, which
is surrounded by the infalling hadronic envelope 
(see Table~\ref{table}).
The mixed phase region shrinks gradually during the PNS collapse
as more and more matter converts from the mixed into the
pure quark phase. On this short time scale of \(\sim 1\) ms,
the SAS remains almost unaffected by this dynamical 
evolution inside the PNS (see Fig.~\ref{fig:hydroplot}).
However, the change in the chemical potentials
and the increasing density during the phase transition 
establish weak equilibrium 
at a lower electron fraction of \(Y_e \leq\) 0.1,
while the mean energy of the trapped \(\nu_e\) 
increases above 200 MeV.

\emph{Shock formation and early shock propagation:}
A subsonic accretion front forms at the interface between the 
hydrostatic pure quark phase and the infalling mixed phase
(thick dashed line Fig.~\ref{fig:hydroplot}).
The accretion front propagates through the mixed phase, 
meets the supersonically infalling hadrons
at the sonic point and turns into an accretion shock (thick dash-dotted line).
The internal energy of the infalling hadronic matter, that is accumulated on 
the compactified core of the PNS, increases
significantly due to the compression in the
strong gravitational field.
This leads to a rapid conversion of hadronic matter into the mixed phase.
As the accreted layers become less dense,
the second accretion shock finally detaches from the mixed phase boundary
and propagates into the pure hadronic phase. This phase was deleptonized by the continued
emission of electron neutrinos after the first neutronization burst.
In this electron-degenerate environment, weak equilibrium is achieved
at an electron fraction $\sim 0.1$. When the second shock runs across
this matter, the electron-degeneracy is lifted by shock-heating
and the weak equilibrium is restored at higher values of the electron
fraction ($Y_e \geq 0.2$).

\emph{Explosion:} As the second shock propagates 
through the outer layers of the PNS,
the ram pressure ahead of the shock decreases rapidly
due to the large decline of density and velocity.
The pressure behind the shock is supported by the accumulation
of the infalling shock-heated matter
and the increasing electron pressure due to the larger \(Y_e\)
in the adjusting weak equilibrium.
Hence, this accretion shock accelerates and turns
into a dynamic shock with positive matter velocities
(thin solid line in Fig.~\ref{fig:hydroplot}).
Up to this point, neutrino transport plays a negligible 
role since neutrinos are trapped.
This changes when the second shock reaches 
the neutrino spheres. A second neutrino burst
of all neutrino flavors is released
(see Fig.~\ref{fig:luminosity-energy}),
dominated by \(\bar{\nu}_e\) due to the 
above-mentioned increase in \(Y_e\). In addition, the more compact
PNS releases (\( \mu / \tau \)) - neutrinos with significantly higher mean
energies as illustrated in Fig.~\ref{fig:luminosity-energy} (c). As soon as the expanding second shock merges with the outer SAS,
the scenario resembles the situation of a
neutrino-driven explosion mechanism (thin dashed line in Fig.~\ref{fig:hydroplot}),
except for the large matter outflow with velocities \(\sim 10^5\) km/s.
Behind the expanding matter, a region with matter inflow develops
due to neutrino cooling (thin dash-dotted line).
The corresponding accretion luminosity from the later fallback explains
the transient increase of the electron neutrino flavor 
luminosities in Fig.~\ref{fig:luminosity-energy} (a) \(\sim\) 340 ms after bounce.
This matter inflow in the cooling region becomes supersonic
and develops another standing accretion shock at the surface of the PNS
at a radius of \(\sim\) 50 km.
The neutrinos emitted from this cooling region
are partly absorbed behind the expanding shock
at a few 100 km radius and support the explosion additionally.
We obtain explosion energies of several \(10^{50}\) erg
(see Table~\ref{table}). 
The neutrino luminosities decrease after 
the onset of the explosion (see Fig.~\ref{fig:luminosity-energy}
at \(\sim\) 350 ms after bounce).

In general, the models with \emph{eos1} and \emph{eos2}
evolve in a qualitatively similar manner.
However, the models with the larger bag constant show
a longer PNS accretion time
before the onset of the phase transition
due to the larger critical density.
This results in a more massive PNS with a 
deeper gravitational potential, so that
the second shock develops larger explosion energies 
(see Table~\ref{table}). 
In comparison to the simulations using \emph{eos1},
the second neutrino burst
appears several 100 ms later and
is found to have a larger peak-luminosity due to 
higher temperatures of the shocked material.
The more massive progenitor stars give an earlier
onset of the phase transition and result in a more massive
PNS. A special case is the dynamical evolution of the PNS 
of the 15 M\(_\odot\) progenitor model using \emph{eos2}.
Almost simultaneously with the formation of the second shock,
the more compact quark core collapses to a black hole.
Still, this does not necessarily exclude an explosion.
Unfortunately, our co-moving coordinate choice does not 
allow us to follow the dynamical evolution beyond the
formation of the apparent horizon \cite{Liebendoerfer:etal:2004}.

The main result of this investigation is a strong 
signature of the formation of quark matter 
in the early postbounce phase of core
collapse supernovae. A second shock forms 
inside the PNS, that affects significantly
the properties of the emitted neutrinos.
For a Galactic core-collapse supernova,
a second neutrino burst should be resolvable 
by the present neutrino detectors.
Unfortunately, the time sequence of the neutrino 
events from SN1987a \cite{Hirata:etal:1988}
was statistically not significant. Anyway, the goal of this first study is to
predict the general effects of the early phase transition to
quark matter during the postbounce phase. To optimally reproduce the observations from SN1987A further analysis and improvements of the EoS would be required. The magnitude and the time delay of the 
second neutrino burst provide correlated information
about the critical density, the EoS in different phases
and the progenitor model.
For low and intermediate mass progenitor models, 
the energy of the second shock
becomes sufficient to drive an explosion
even in spherical symmetry.
The explosion is powered by the accretion of matter into 
the deeper gravitational potential of the more compact PNS,
if quark matter is present in the core.
The ejecta contain neutron-rich material that expands on a fast timescale
and should be investigated as a possible site for the r-process.
With respect to the remnant, the narrow range of PNS masses
found in Table~\ref{table}
may provide an explanation for the clustering of the observed 
neutron star masses (gravitational) around 1.4 M\(_\odot\) 
(see e.g.~\cite{LattimerPrakash:2004}).
The discussed direct black hole formation at the phase transition 
could be investigated further in light of the observed 
connection between supernovae and \(\gamma\)-ray bursts \cite{Piran:2005}.
For other progenitor models, a delayed collapse to a black hole may
occur after the explosion during the PNS cooling phase.

The presented analysis should be complemented by 
multi-dimensional simulations, to explore the impact of
known fluid instabilities that can not be treated in spherical symmetry.
Another interesting scenario would be a weak neutrino driven explosion,
followed by a fallback-induced QCD phase transition.
Since the QCD phase diagram shows a large variety
of color-superconducting phases \cite{Ruester:etal:2005,Blaschke:etal:2005,Sandin:2007zr},
a more sophisticated quark matter EoS should be adopted.
This could lead to a second phase transition within the quark core of the PNS
and would be an interesting extension of the present study.

This work has been supported by the
Swiss National Science Foundation under the grant numbers 
PP002-106627/1 and PP200020-105328/1,
the Helmholtz Research School for Quark Matter Studies,
the Italian National Institute for Nuclear Physics,
the Frankfurt Institute for Advanced Studies
and
by the ESF CompStar program.
A.M. is supported at the Oak Ridge National Laboratory, which is managed by
UT-Battelle, LLC for the U.S. Department of Energy under contract
DE-AC05-00OR22725.
We would like to thank for stimulating discussions with 
D.~Blaschke, A.~Drago, G.~Mart\'inez-Pinedo and S.~C.~Whitehouse.

\bibliographystyle{apsrev}

\begin{thebibliography}{31}
\expandafter\ifx\csname natexlab\endcsname\relax\def\natexlab#1{#1}\fi
\expandafter\ifx\csname bibnamefont\endcsname\relax
  \def\bibnamefont#1{#1}\fi
\expandafter\ifx\csname bibfnamefont\endcsname\relax
  \def\bibfnamefont#1{#1}\fi
\expandafter\ifx\csname citenamefont\endcsname\relax
  \def\citenamefont#1{#1}\fi
\expandafter\ifx\csname url\endcsname\relax
  \def\url#1{\texttt{#1}}\fi
\expandafter\ifx\csname urlprefix\endcsname\relax\def\urlprefix{URL }\fi
\providecommand{\bibinfo}[2]{#2}
\providecommand{\eprint}[2][]{\url{#2}}

\bibitem[{\citenamefont{{Weber}}(2005)}]{Weber:2005}
\bibinfo{author}{\bibfnamefont{F.}~\bibnamefont{{Weber}}},
  \bibinfo{journal}{Progress in Particle and Nuclear Physics}
  \textbf{\bibinfo{volume}{54}}, \bibinfo{pages}{193} (\bibinfo{year}{2005}).

\bibitem[{\citenamefont{Pons et~al.}(2001)\citenamefont{Pons, Steiner, Prakash
  and Lattimer}}]{Pons:2001ar}
\bibinfo{author}{\bibfnamefont{J.~A.} \bibnamefont{Pons}},
  \bibinfo{author}{\bibfnamefont{A.~W.} \bibnamefont{Steiner}},
  \bibinfo{author}{\bibfnamefont{M.}~\bibnamefont{Prakash}} \bibnamefont{and}
  \bibinfo{author}{\bibfnamefont{J.~M.} \bibnamefont{Lattimer}},
  \bibinfo{journal}{Phys. Rev. Lett.} \textbf{\bibinfo{volume}{86}},
  \bibinfo{pages}{5223} (\bibinfo{year}{2001}).

\bibitem[{\citenamefont{Lin et~al.}(2006)\citenamefont{Lin, Cheng, Chu, and
  Suen}}]{Lin:2005zda}
\bibinfo{author}{\bibfnamefont{L.-M.} \bibnamefont{Lin}},
  \bibinfo{author}{\bibfnamefont{K.~S.} \bibnamefont{Cheng}},
  \bibinfo{author}{\bibfnamefont{M.~C.} \bibnamefont{Chu}} \bibnamefont{and}
  \bibinfo{author}{\bibfnamefont{W.~M.} \bibnamefont{Suen}},
  \bibinfo{journal}{Astrophys. J.} \textbf{\bibinfo{volume}{639}},
  \bibinfo{pages}{382} (\bibinfo{year}{2006}).

\bibitem[{\citenamefont{Abdikamalov et~al.}(2008)\citenamefont{Abdikamalov,
  Dimmelmeier, Rezzolla and Miller}}]{Abdikamalov:2008df}
\bibinfo{author}{\bibfnamefont{E.~B.} \bibnamefont{Abdikamalov}},
  \bibinfo{author}{\bibfnamefont{H.}~\bibnamefont{Dimmelmeier}},
  \bibinfo{author}{\bibfnamefont{L.}~\bibnamefont{Rezzolla}} \bibnamefont{and}
  \bibinfo{author}{\bibfnamefont{J.~C.} \bibnamefont{Miller}}
  (\bibinfo{year}{2008}), \eprint{astro-ph/0806.1700}.

\bibitem[{\citenamefont{Berezhiani et~al.}(2003)\citenamefont{Berezhiani,
  Bombaci, Drago, Frontera and Lavagno}}]{Berezhiani:2002ks}
\bibinfo{author}{\bibfnamefont{Z.}~\bibnamefont{Berezhiani}},
  \bibinfo{author}{\bibfnamefont{I.}~\bibnamefont{Bombaci}},
  \bibinfo{author}{\bibfnamefont{A.}~\bibnamefont{Drago}},
  \bibinfo{author}{\bibfnamefont{F.}~\bibnamefont{Frontera}} \bibnamefont{and}
  \bibinfo{author}{\bibfnamefont{A.}~\bibnamefont{Lavagno}},
  \bibinfo{journal}{Astrophys. J.} \textbf{\bibinfo{volume}{586}},
  \bibinfo{pages}{1250} (\bibinfo{year}{2003}).

\bibitem[{\citenamefont{{Takahara} and
  {Sato}}(1988{\natexlab{a}})}]{TakaharaSato:1988a}
\bibinfo{author}{\bibfnamefont{M.}~\bibnamefont{{Takahara}}} \bibnamefont{and}
  \bibinfo{author}{\bibfnamefont{K.}~\bibnamefont{{Sato}}},
  \bibinfo{journal}{\apj} \textbf{\bibinfo{volume}{335}}, \bibinfo{pages}{301}
  (\bibinfo{year}{1988}{\natexlab{a}}).

\bibitem[{\citenamefont{{Takahara} and
  {Sato}}(1988{\natexlab{b}})}]{TakaharaSato:1988b}
\bibinfo{author}{\bibfnamefont{M.}~\bibnamefont{{Takahara}}} \bibnamefont{and}
  \bibinfo{author}{\bibfnamefont{K.}~\bibnamefont{{Sato}}},
  \bibinfo{journal}{Progress of Theoretical Physics}
  \textbf{\bibinfo{volume}{80}}, \bibinfo{pages}{861}
  (\bibinfo{year}{1988}{\natexlab{b}}).

\bibitem[{\citenamefont{Gentile et~al.}(1993)\citenamefont{Gentile,
  Aufderheide, Mathews, Swesty and Fuller}}]{Gentile:1993ma}
\bibinfo{author}{\bibfnamefont{N.~A.} \bibnamefont{Gentile}},
  \bibinfo{author}{\bibfnamefont{M.~B.} \bibnamefont{Aufderheide}},
  \bibinfo{author}{\bibfnamefont{G.~J.} \bibnamefont{Mathews}},
  \bibinfo{author}{\bibfnamefont{F.~D.} \bibnamefont{Swesty}},
  \bibnamefont{and} \bibinfo{author}{\bibfnamefont{G.~M.}
  \bibnamefont{Fuller}}, \bibinfo{journal}{Astrophys. J.}
  \textbf{\bibinfo{volume}{414}}, \bibinfo{pages}{701} (\bibinfo{year}{1993}).

\bibitem[{\citenamefont{Drago and Tambini}(1999)}]{Drago:1997tn}
\bibinfo{author}{\bibfnamefont{A.}~\bibnamefont{Drago}} \bibnamefont{and}
  \bibinfo{author}{\bibfnamefont{U.}~\bibnamefont{Tambini}},
  \bibinfo{journal}{J. Phys.} \textbf{\bibinfo{volume}{G25}},
  \bibinfo{pages}{971} (\bibinfo{year}{1999}).

\bibitem[{\citenamefont{Yasutake et~al.}(2007)\citenamefont{Yasutake, Kotake,
  Hashimoto, and Yamada}}]{Yasutake:2007st}
\bibinfo{author}{\bibfnamefont{N.}~\bibnamefont{Yasutake}},
  \bibinfo{author}{\bibfnamefont{K.}~\bibnamefont{Kotake}},
  \bibinfo{author}{\bibfnamefont{M.-a.} \bibnamefont{Hashimoto}},
  \bibnamefont{and} \bibinfo{author}{\bibfnamefont{S.}~\bibnamefont{Yamada}},
  \bibinfo{journal}{Phys. Rev.} \textbf{\bibinfo{volume}{D75}},
  \bibinfo{pages}{084012} (\bibinfo{year}{2007}).

\bibitem[{\citenamefont{Nakazato et~al.}(2008)\citenamefont{Nakazato,
  Sumiyoshi and Yamada}}]{Nakazato:2008su}
\bibinfo{author}{\bibfnamefont{K.}~\bibnamefont{Nakazato}},
  \bibinfo{author}{\bibfnamefont{K.}~\bibnamefont{Sumiyoshi}},
  \bibnamefont{and} \bibinfo{author}{\bibfnamefont{S.}~\bibnamefont{Yamada}},
  \bibinfo{journal}{Phys. Rev.} \textbf{\bibinfo{volume}{D77}},
  \bibinfo{pages}{103006} (\bibinfo{year}{2008}).

\bibitem[{\citenamefont{{Bethe} and {Wilson}}(1985)}]{BetheWilson:1985}
\bibinfo{author}{\bibfnamefont{H.~A.} \bibnamefont{{Bethe}}} \bibnamefont{and}
  \bibinfo{author}{\bibfnamefont{J.~R.} \bibnamefont{{Wilson}}},
  \bibinfo{journal}{\apj} \textbf{\bibinfo{volume}{295}}, \bibinfo{pages}{14}
  (\bibinfo{year}{1985}).

\bibitem[{\citenamefont{{LeBlanc} and {Wilson}}(1970)}]{LeBlankWilson:1970}
\bibinfo{author}{\bibfnamefont{J.~M.} \bibnamefont{{LeBlanc}}}
  \bibnamefont{and} \bibinfo{author}{\bibfnamefont{J.~R.}
  \bibnamefont{{Wilson}}}, \bibinfo{journal}{\apj}
  \textbf{\bibinfo{volume}{161}}, \bibinfo{pages}{541} (\bibinfo{year}{1970}).

\bibitem[{\citenamefont{{Bisnovatyi-Kogan}}(1971)}]{Bisnovatyi-Kogan:1971}
\bibinfo{author}{\bibfnamefont{G.~S.} \bibnamefont{{Bisnovatyi-Kogan}}},
  \bibinfo{journal}{Soviet Astronomy} \textbf{\bibinfo{volume}{14}},
  \bibinfo{pages}{652} (\bibinfo{year}{1971}).

\bibitem[{\citenamefont{{Burrows} et~al.}(2006)\citenamefont{{Burrows},
  {Livne}, {Dessart}, {Ott} and {Murphy}}}]{Burrows:etal:2006}
\bibinfo{author}{\bibfnamefont{A.}~\bibnamefont{{Burrows}}},
  \bibinfo{author}{\bibfnamefont{E.}~\bibnamefont{{Livne}}},
  \bibinfo{author}{\bibfnamefont{L.}~\bibnamefont{{Dessart}}},
  \bibinfo{author}{\bibfnamefont{C.~D.} \bibnamefont{{Ott}}} \bibnamefont{and}
  \bibinfo{author}{\bibfnamefont{J.}~\bibnamefont{{Murphy}}},
  \bibinfo{journal}{New Astronomy Review} \textbf{\bibinfo{volume}{50}},
  \bibinfo{pages}{487} (\bibinfo{year}{2006}).

\bibitem[{\citenamefont{Schertler et~al.}(2000)\citenamefont{Schertler,
  Greiner, Schaffner-Bielich, and Thoma}}]{Schertler:2000xq}
\bibinfo{author}{\bibfnamefont{K.}~\bibnamefont{Schertler}},
  \bibinfo{author}{\bibfnamefont{C.}~\bibnamefont{Greiner}},
  \bibinfo{author}{\bibfnamefont{J.}~\bibnamefont{Schaffner-Bielich}},
  \bibnamefont{and} \bibinfo{author}{\bibfnamefont{M.~H.} \bibnamefont{Thoma}},
  \bibinfo{journal}{Nucl. Phys.} \textbf{\bibinfo{volume}{A677}},
  \bibinfo{pages}{463} (\bibinfo{year}{2000}), \eprint{astro-ph/0001467}.


\bibitem[{\citenamefont{Eidelman et~al.}(2004)}]{Eidelman:2004wy}
\bibinfo{author}{\bibfnamefont{S.}~\bibnamefont{Eidelman}} \bibnamefont{et~al.}
  (\bibinfo{collaboration}{Particle Data Group}), \bibinfo{journal}{Phys.
  Lett.} \textbf{\bibinfo{volume}{B592}}, \bibinfo{pages}{1}
  (\bibinfo{year}{2004}).

\bibitem[{\citenamefont{Shen et~al.}(1998)\citenamefont{Shen, Toki, Oyamatsu,
  and Sumiyoshi}}]{Shen:1998gq}
\bibinfo{author}{\bibfnamefont{H.}~\bibnamefont{Shen}},
  \bibinfo{author}{\bibfnamefont{H.}~\bibnamefont{Toki}},
  \bibinfo{author}{\bibfnamefont{K.}~\bibnamefont{Oyamatsu}} \bibnamefont{and}
  \bibinfo{author}{\bibfnamefont{K.}~\bibnamefont{Sumiyoshi}},
  \bibinfo{journal}{Nucl. Phys.} \textbf{\bibinfo{volume}{A637}},
  \bibinfo{pages}{435} (\bibinfo{year}{1998}).

\bibitem[{\citenamefont{{Landau} and {Lifshitz}}(1969)}]{1969stph.book.....L}
\bibinfo{author}{\bibfnamefont{L.~D.} \bibnamefont{{Landau}}} \bibnamefont{and}
  \bibinfo{author}{\bibfnamefont{E.~M.} \bibnamefont{{Lifshitz}}},
  \emph{\bibinfo{title}{{Statistical physics. Pt.1}}}
  (\bibinfo{publisher}{Course of theoretical physics - Pergamon International
  Library of Science, Technology, Engineering and Social Studies, Oxford:
  Pergamon Press, and Reading: Addison-Wesley, |c1969, 2nd rev.~- enlarg.ed.},
  \bibinfo{year}{1969}).



\bibitem[{\citenamefont{Voskresensky et~al.}(2003)\citenamefont{Voskresensky,
  Yasuhira and Tatsumi}}]{Voskresensky:2002hu}
\bibinfo{author}{\bibfnamefont{D.~N.} \bibnamefont{Voskresensky}},
  \bibinfo{author}{\bibfnamefont{M.}~\bibnamefont{Yasuhira}} \bibnamefont{and}
  \bibinfo{author}{\bibfnamefont{T.}~\bibnamefont{Tatsumi}},
  \bibinfo{journal}{Nucl. Phys.} \textbf{\bibinfo{volume}{A723}},
  \bibinfo{pages}{291} (\bibinfo{year}{2003}).

\bibitem[{\citenamefont{{Lattimer} and {Prakash}}(2004)}]{LattimerPrakash:2004}
\bibinfo{author}{\bibfnamefont{J.~M.} \bibnamefont{{Lattimer}}}
  \bibnamefont{and}
  \bibinfo{author}{\bibfnamefont{M.}~\bibnamefont{{Prakash}}},
  \bibinfo{journal}{Science} \textbf{\bibinfo{volume}{304}},
  \bibinfo{pages}{536} (\bibinfo{year}{2004}).

\bibitem[{\citenamefont{Alford et~al.}(2007)}]{Alford:2006vz}
\bibinfo{author}{\bibfnamefont{M.}~\bibnamefont{Alford}} \bibnamefont{et~al.},
  \bibinfo{journal}{Nature} \textbf{\bibinfo{volume}{445}}, \bibinfo{pages}{E7}
  (\bibinfo{year}{2007}).

\bibitem[{\citenamefont{{Mezzacappa} and
  {Bruenn}}(1993)}]{MezzacappaBruenn:1993c}
\bibinfo{author}{\bibfnamefont{A.}~\bibnamefont{{Mezzacappa}}}
  \bibnamefont{and} \bibinfo{author}{\bibfnamefont{S.~W.}
  \bibnamefont{{Bruenn}}}, \bibinfo{journal}{\apj}
  \textbf{\bibinfo{volume}{410}}, \bibinfo{pages}{740} (\bibinfo{year}{1993}).

\bibitem[{\citenamefont{{Mezzacappa} and
  {Messer}}(1999)}]{MezzacappaMesser:1999}
\bibinfo{author}{\bibfnamefont{A.}~\bibnamefont{{Mezzacappa}}}
  \bibnamefont{and} \bibinfo{author}{\bibfnamefont{O.~E.~B.}
  \bibnamefont{{Messer}}}, \bibinfo{journal}{Journal of Computational and
  Applied Mathematics} \textbf{\bibinfo{volume}{109}}, \bibinfo{pages}{281}
  (\bibinfo{year}{1999}).

\bibitem[{\citenamefont{{Liebend{\"o}rfer}
  et~al.}(2004)\citenamefont{{Liebend{\"o}rfer}, {Messer}, {Mezzacappa},
  {Bruenn}, {Cardall} and {Thielemann}}}]{Liebendoerfer:etal:2004}
\bibinfo{author}{\bibfnamefont{M.}~\bibnamefont{{Liebend{\"o}rfer}}},
  \bibinfo{author}{\bibfnamefont{O.~E.~B.} \bibnamefont{{Messer}}},
  \bibinfo{author}{\bibfnamefont{A.}~\bibnamefont{{Mezzacappa}}},
  \bibinfo{author}{\bibfnamefont{S.~W.} \bibnamefont{{Bruenn}}},
  \bibinfo{author}{\bibfnamefont{C.~Y.} \bibnamefont{{Cardall}}},
  \bibnamefont{and} \bibinfo{author}{\bibfnamefont{F.-K.}
  \bibnamefont{{Thielemann}}}, \bibinfo{journal}{\apj Supplement}
  \textbf{\bibinfo{volume}{150}}, \bibinfo{pages}{263} (\bibinfo{year}{2004}).

\bibitem[{\citenamefont{{Bruenn}}(1985)}]{Bruenn:1985}
\bibinfo{author}{\bibfnamefont{S.~W.} \bibnamefont{{Bruenn}}},
  \bibinfo{journal}{\apj Supplement} \textbf{\bibinfo{volume}{58}},
  \bibinfo{pages}{771} (\bibinfo{year}{1985}).

\bibitem[{\citenamefont{{Woosley} et~al.}(2002)\citenamefont{{Woosley},
  {Heger} and {Weaver}}}]{Woosley:etal:2002}
\bibinfo{author}{\bibfnamefont{S.~E.} \bibnamefont{{Woosley}}},
  \bibinfo{author}{\bibfnamefont{A.}~\bibnamefont{{Heger}}} \bibnamefont{and}
  \bibinfo{author}{\bibfnamefont{T.~A.} \bibnamefont{{Weaver}}},
  \bibinfo{journal}{Reviews of Modern Physics} \textbf{\bibinfo{volume}{74}},
  \bibinfo{pages}{1015} (\bibinfo{year}{2002}).

\bibitem[{\citenamefont{{Kitaura} et~al.}(2006)\citenamefont{{Kitaura},
  {Janka} and {Hillebrandt}}}]{Kitaura:etal:2006}
\bibinfo{author}{\bibfnamefont{F.~S.} \bibnamefont{{Kitaura}}},
  \bibinfo{author}{\bibfnamefont{H.-T.} \bibnamefont{{Janka}}},
  \bibnamefont{and}
  \bibinfo{author}{\bibfnamefont{W.}~\bibnamefont{{Hillebrandt}}},
  \bibinfo{journal}{A\&A} \textbf{\bibinfo{volume}{450}}, \bibinfo{pages}{345}
  (\bibinfo{year}{2006}).

\bibitem[{\citenamefont{Steiner et~al.}(2000)\citenamefont{Steiner, Prakash
  and Lattimer}}]{Steiner:2000bi}
\bibinfo{author}{\bibfnamefont{A.}~\bibnamefont{Steiner}},
  \bibinfo{author}{\bibfnamefont{M.}~\bibnamefont{Prakash}} \bibnamefont{and}
  \bibinfo{author}{\bibfnamefont{J.~M.} \bibnamefont{Lattimer}},
  \bibinfo{journal}{Phys. Lett.} \textbf{\bibinfo{volume}{B486}},
  \bibinfo{pages}{239} (\bibinfo{year}{2000}).

\bibitem[{\citenamefont{{Hirata} et~al.}(1988)\citenamefont{{Hirata}, {Kajita},
  {Koshiba}, {Nakahata}, {Oyama}, {Sato}, {Suzuki}, {Takita}, {Totsuka},
  {Kifune} et~al.}}]{Hirata:etal:1988}
\bibinfo{author}{\bibfnamefont{K.~S.} \bibnamefont{{Hirata}}},
  \bibinfo{author}{\bibfnamefont{T.}~\bibnamefont{{Kajita}}},
  \bibinfo{author}{\bibfnamefont{M.}~\bibnamefont{{Koshiba}}},
  \bibinfo{author}{\bibfnamefont{M.}~\bibnamefont{{Nakahata}}},
  \bibinfo{author}{\bibfnamefont{Y.}~\bibnamefont{{Oyama}}},
  \bibinfo{author}{\bibfnamefont{N.}~\bibnamefont{{Sato}}},
  \bibinfo{author}{\bibfnamefont{A.}~\bibnamefont{{Suzuki}}},
  \bibinfo{author}{\bibfnamefont{M.}~\bibnamefont{{Takita}}},
  \bibinfo{author}{\bibfnamefont{Y.}~\bibnamefont{{Totsuka}}},
  \bibinfo{author}{\bibfnamefont{T.}~\bibnamefont{{Kifune}}},
  \bibnamefont{et~al.}, \bibinfo{journal}{\prd} \textbf{\bibinfo{volume}{38}},
  \bibinfo{pages}{448} (\bibinfo{year}{1988}).

\bibitem[{\citenamefont{{Piran}}(2005)}]{Piran:2005}
\bibinfo{author}{\bibfnamefont{T.}~\bibnamefont{{Piran}}},
  \bibinfo{journal}{Reviews of Modern Physics} \textbf{\bibinfo{volume}{76}},
  \bibinfo{pages}{1143} (\bibinfo{year}{2005}).

\bibitem[{\citenamefont{{R{\"u}ster} et~al.}(2005)\citenamefont{{R{\"u}ster},
  {Werth}, {Buballa}, {Shovkovy} and {Rischke}}}]{Ruester:etal:2005}
\bibinfo{author}{\bibfnamefont{S.~B.} \bibnamefont{{R{\"u}ster}}},
  \bibinfo{author}{\bibfnamefont{V.}~\bibnamefont{{Werth}}},
  \bibinfo{author}{\bibfnamefont{M.}~\bibnamefont{{Buballa}}},
  \bibinfo{author}{\bibfnamefont{I.~A.} \bibnamefont{{Shovkovy}}},
  \bibnamefont{and} \bibinfo{author}{\bibfnamefont{D.~H.}
  \bibnamefont{{Rischke}}}, \bibinfo{journal}{\prd}
  \textbf{\bibinfo{volume}{72}}, \bibinfo{pages}{034004}
  (\bibinfo{year}{2005}).

\bibitem[{\citenamefont{{Blaschke} et~al.}(2005)\citenamefont{{Blaschke},
  {Fredriksson}, {Grigorian}, {{\"O}zta{\c s}} and
  {Sandin}}}]{Blaschke:etal:2005}
\bibinfo{author}{\bibfnamefont{D.}~\bibnamefont{{Blaschke}}},
  \bibinfo{author}{\bibfnamefont{S.}~\bibnamefont{{Fredriksson}}},
  \bibinfo{author}{\bibfnamefont{H.}~\bibnamefont{{Grigorian}}},
  \bibinfo{author}{\bibfnamefont{A.~M.} \bibnamefont{{{\"O}zta{\c s}}}},
  \bibnamefont{and} \bibinfo{author}{\bibfnamefont{F.}~\bibnamefont{{Sandin}}},
  \bibinfo{journal}{\prd} \textbf{\bibinfo{volume}{72}},
  \bibinfo{pages}{065020} (\bibinfo{year}{2005}).

\bibitem[{\citenamefont{Sandin and Blaschke}(2007)}]{Sandin:2007zr}
\bibinfo{author}{\bibfnamefont{F.}~\bibnamefont{Sandin}} \bibnamefont{and}
  \bibinfo{author}{\bibfnamefont{D.}~\bibnamefont{Blaschke}},
  \bibinfo{journal}{Phys. Rev.} \textbf{\bibinfo{volume}{D75}},
  \bibinfo{pages}{125013} (\bibinfo{year}{2007}), \eprint{astro-ph/0701772}.


\end{thebibliography}

\end{document}